\documentclass[10pt,preprint]{aastex}
\defcitealias{car99}{CM99}
\defcitealias{and03}{AK03}

\begin{document}
\title{Proper Motion of the Crab Pulsar Revisited}
\author{C.-Y. Ng and Roger W. Romani}
\affil{Department of Physics, Stanford University, Stanford, CA 94305}
\email{ncy@astro.stanford.edu, rwr@astro.stanford.edu}

\begin{abstract}
It has been suggested that the Crab pulsar's proper motion is well 
aligned with the symmetry axis of the pulsar wind nebula.
We have re-visited this question,
examining over 6 years of F547M WFPC2 chip 3 images to obtain a best-fit 
value of $\mu_\ast = 14.9 \pm 0.8$mas/yr at PA $278\arcdeg \pm 3\arcdeg$.
At $26\arcdeg\pm 3\arcdeg$ to the nebula axis,
this substantially relaxes constraints on the birth kick of this pulsar.
Such misalignment allows the momentum to be imparted over $\sim$1s timescales.
\end{abstract}

\section{Introduction}
Pulsars are fast-moving objects with space velocities up to an order of 
magnitude larger than their progenitors.  It has long been suggested that 
this is the result of a momentum kick at birth.  One way to probe the kick 
physics is to compare the pulsar's velocity and spin axis orientations.
The kick timescale is constrained by the alignment between the two vectors.
For the Crab pulsar, stunning images from the Chandra X-ray Observatory 
reveal the torus-like pulsar wind nebula.  The symmetry axis indicates the 3D 
orientation of the pulsar spin axis, and more interestingly, suggests an 
alignment with the proper motion.  \citet{ng04} fitted the pulsar wind 
torus and derived a quantitative measurement of the spin axis at position 
angle (PA) $124\fdg0$ or $304\fdg0 \pm0\fdg1$.
In contrast, the pulsar's proper motion is not as well measured.
Previous works by \citet{wyc77} and \citet{car99} \citepalias[hereafter][]{car99} 
have large uncertainties.  The former examined photographic plates over 
77 years and obtained the result of $13\pm4$ mas/yr, at PA $293\arcdeg\pm18\arcdeg$.
\citetalias{car99} performed relative astrometry on the Wide Field 
Planetary Camera 2 (WFPC2) images from the Hubble Space Telescope (HST) 
and found a proper motion of $\mu_\ast=18\pm3$ mas/yr at 
$292\arcdeg\pm10\arcdeg$ using dataset spanning 1.9 years.

The high angular resolution cameras on board the HST make it possible to 
determine the pulsar proper motion in a short period of time.
In the past few years, many new observations of the Crab pulsar have been 
collected with HST. The HST archive in fact contains WFPC2 Crab images
spanning some 7 years, and new Advanced Camera for Surveys (ACS)
images are now being taken. Also, recent studies have improved our
understanding of the WFPC2 camera's geometric distortions 
\citep[][hereafter \citetalias{and03}]{and03}.  Thus, with a longer 
timebase and improved calibration, this study aims to update \citetalias{car99}'s
estimate of the Crab pulsar's proper motion.

\section{The Datasets}

Using HST/WFPC2 images, relative astrometry can be made to an accuracy of 
$0\farcs005$ for objects on the same chip \citep[section 5.4]{bag02}.
We have searched the Multi-mission Archive at Space Telescope (MAST) and 
found more than 70 WFPC2 observations of the pulsar from 1994 to 2001,
making this the best instrument for the proper motion study. These
images were taken using a variety of optical filters.  However, as 
plate scales and geometric distortion vary slightly for different filters, 
we restrict our analysis to a single one.  The F547M filter is the best choice,
having the most observations and the longest time span of 7 years.
Table~\ref{tab1} lists all the observations with this filter; epochs with similar 
pointing parameters are grouped.  Obs~8a, which is observed with the 
POLQ filter, is also listed in the table as it is used in \citetalias{car99}.
We have included the first ACS observation in the table as well.

\begin{deluxetable}{cccccccccc}
\tabletypesize{\scriptsize}
\tablecaption{WFPC2 observations of the Crab pulsar with the F547M filter.\label{tab1}}
\tablewidth{0pt}
\tablehead{
\colhead{Obs.} & \colhead{Date} & \colhead{Exp. (s)} & \colhead{Detector} & \colhead{\# Ref. Stars} & \colhead{Roll Angle} & \colhead{Chip x} & \colhead{Chip y} & \colhead{Group}
}
\startdata
1 & 1994-03-09 & 2000 & WF2 & 11 & -116.8 & 463 & 160 & \\ \tableline
2 & 1995-01-06 & 1600 & WF3 & 8 & -51.2 & 163 & 130 & 1 \\
3 & 1995-01-06 & 2000 & WF3 & 7 & -51.2 & 151 & 118 & 1 \\ \tableline
4 & 1995-08-14 & 2000 & PC & 5 & -48.7 & 470 & 371 & \\
5 & 1995-11-05 & 2000 & PC & 2 & -25.3 & 340 & 300 & \\
6 & 1995-12-29 & 2000 & PC & 4 & 128.7 & 297 & 600 & 2 \\
7 & 1996-01-20 & 2000 & PC & 4 & 128.7 & 297 & 601 & 2 \\
8 & 1996-01-26 & 2000 & PC & 4 & 128.7 & 297 & 603 & 2 \\
8a & 1996-01-26 & 2000 & WF2 & 13 & -153.3 & 495 & 546 &  \\
9 & 1996-02-01 & 2000 & PC & 4 & 128.7 & 297 & 601 & 2 \\
10 & 1996-02-22 & 2000 & PC & 4 & 128.7 & 297 & 602 & 2 \\
11 & 1996-04-16 & 2000 & PC & 4 & 128.9 & 266 & 599 & 2 \\ \tableline
12 & 2000-02-06 & 2200 & WF3 & 15 & -47.3 & 373 & 273 & 3 \\
13 & 2000-02-15 & 2200 & WF3 & 15 & -47.3 & 372 & 273 & 3 \\
14 & 2000-02-26 & 2200 & WF3 & 15 & -47.3 & 371 & 271 & 3 \\
15 & 2000-03-07 & 2200 & WF3 & 15 & -47.3 & 372 & 272 & 3 \\
16 & 2000-03-17 & 2200 & WF3 & 15 & -47.3 & 373 & 273 & 3 \\ \tableline
17 & 2000-08-07 & 2300 & WF4 & 6 & -136.7 & 63 & 136 & 4 \\
18 & 2000-08-18 & 2400 & WF4 & 6 & -136.7 & 63 & 136 & 4 \\
19 & 2000-08-29 & 2400 & WF4 & 6 & -136.7 & 64 & 136 & 4 \\ \tableline
20 & 2000-09-09 & 1837 & WF3 & 8 & 132.7 & 217 & 342 & 5 \\
21 & 2000-09-21 & 2071 & WF3 & 8 & 132.7 & 219 & 342 & 5 \\
22 & 2000-10-01 & 1854 & WF3 & 8 & 132.7 & 217 & 342 & 5 \\
23 & 2000-10-12 & 2400 & WF3 & 8 & 132.7 & 217 & 341 & 5 \\
24 & 2000-10-23 & 2007 & WF3 & 8 & 132.7 & 217 & 342 & 5 \\
25 & 2000-11-15 & 1927 & WF3 & 8 & 132.7 & 217 & 342 & 5 \\
26 & 2000-11-25 & 2300 & WF3 & 8 & 132.7 & 226 & 334 & 5 \\
27 & 2000-12-06 & 2100 & WF3 & 8 & 132.7 & 217 & 340 & 5 \\ \tableline
28 & 2000-12-18 & 2100 & WF3 & 15 & -47.3 & 372 & 274 & 6 \\
29 & 2000-12-28 & 2100 & WF3 & 15 & -47.3 & 372 & 273 & 6 \\
30 & 2001-01-09 & 2400 & WF3 & 15 & -47.3 & 378 & 275 & 6 \\
31 & 2001-01-19 & 2400 & WF3 & 15 & -47.3 & 388 & 276 & 6 \\
32 & 2001-01-30 & 2400 & WF3 & 15 & -47.3 & 388 & 276 & 6 \\
33 & 2001-02-10 & 2400 & WF3 & 15 & -47.3 & 384 & 278 & 6 \\
34 & 2001-02-21 & 2400 & WF3 & 15 & -47.3 & 388 & 276 & 6 \\
35 & 2001-03-04 & 2400 & WF3 & 15 & -47.3 & 389 & 275 & 6 \\
36 & 2001-03-15 & 2400 & WF3 & 15 & -47.3 & 390 & 275 & 6 \\
37 & 2001-03-26 & 2400 & WF3 & 15 & -47.3 & 390 & 275 & 6 \\
38 & 2001-04-06 & 2400 & WF3 & 15 & -47.3 & 390 & 275 & 6 \\
39 & 2001-04-17 & 2400 & WF3 & 15 & -47.3 & 390 & 276 & 6 \\ \tableline
40 & 2001-04-19 & 2400 & WF2 & 6 & -137.6 & 133 & 41 & \\ \tableline
41 & 2003-08-08 & 2200 & ACS & 14 & -92.6 & & &
\enddata 
\end{deluxetable}
\section{Data Reduction}

To start, we attempted to reproduce the \citetalias{car99} results.
We used the same images (obs~1, 4 \& 8a) and the same reference stars
and attempted to follow their analysis as closely as possible.
The data retrieved from the MAST are processed with the On-The-Fly 
Reprocessing (OTFR) system to ensure up-to-date calibration.
The task \emph{crrej} is employed to remove the cosmic rays and 
co-add images for each epoch.  Geometric distortion is then 
corrected by remapping the images, using the task \emph{wmosaic} 
in the \emph{STSDAS} package.  Coordinates of the pulsar and 
the four reference stars are obtained from a 2D Gaussian fit with
the task \emph{fitpsf} in \emph{IRAF}.  Finally, the best-fit 
positions are rotated according to the roll angles in image headers 
and the frames are offset to align the reference stars with obs~1.

By following \citetalias{car99}'s procedure in this way, we obtain
\[ \mu''_\alpha \equiv \mu_\alpha \cos \delta = -11\pm 15\;\mathrm{mas/yr,} 
\;\; \mu_\delta = 2.0\pm 90\;\mathrm{mas/yr,} \]
i.e. no significant detection of the proper motion.
This is obviously inconsistent with \citetalias{car99}.
The extremely large uncertainty is caused by star~1,
which suffers a relative shift of 2 pixels from obs~1 to obs~8a.
Comparison with other field stars shows that the shift is systematic and
due to large changes in the field
distortion caused by the extra polarizer filter POLQ used in obs~8a. This
underscores the fact that different filters are incompatible and should 
not be compared directly without filter-specific distortion maps.
If we exclude frame~8a and fit only the first two images,
the star residuals after frame alignment are 7.6 mas. With four reference
stars, the formal statistical uncertainty on frame alignment is
$\sim$ half as large, but as systematic distortions dominate the alignment
error, the pulsar astrometry is no better than the 7.6 mas level.  
We feel that this is a more realistic assessment of the astrometric
errors in these data than the \citetalias{car99} values. This gives
$\mu''_\alpha  = -10.2\pm 4.9\;\mathrm{mas/yr,} 
~\mu_\delta = 2.0\pm 5.6\;\mathrm{mas/yr}$ a rather low significance
detection of the Crab proper motion.
In any case, we see no way to avoid large errors in star astrometry
and frame registration with simple Gaussian fits to only a few star
images.

\subsection{Our Approach}

Since many F547M frames are now available, we can make a more
careful assessment of the proper motion, restricting to a single filter
configuration and examining similar telescope pointings. Also, we should
avoid performing distortion corrections with the task \emph{wmosaic},
which degrades the resolution by re-sampling the image. Instead,
we followed an approach similar to \citet*{kap02} that involves no 
re-sampling of the data.  Star coordinates are first measured on the chip
reference frame. These pixel positions are then corrected for geometric 
distortion, followed by the standard transformation for frame alignment.

\begin{figure}[!ht]
\plotone{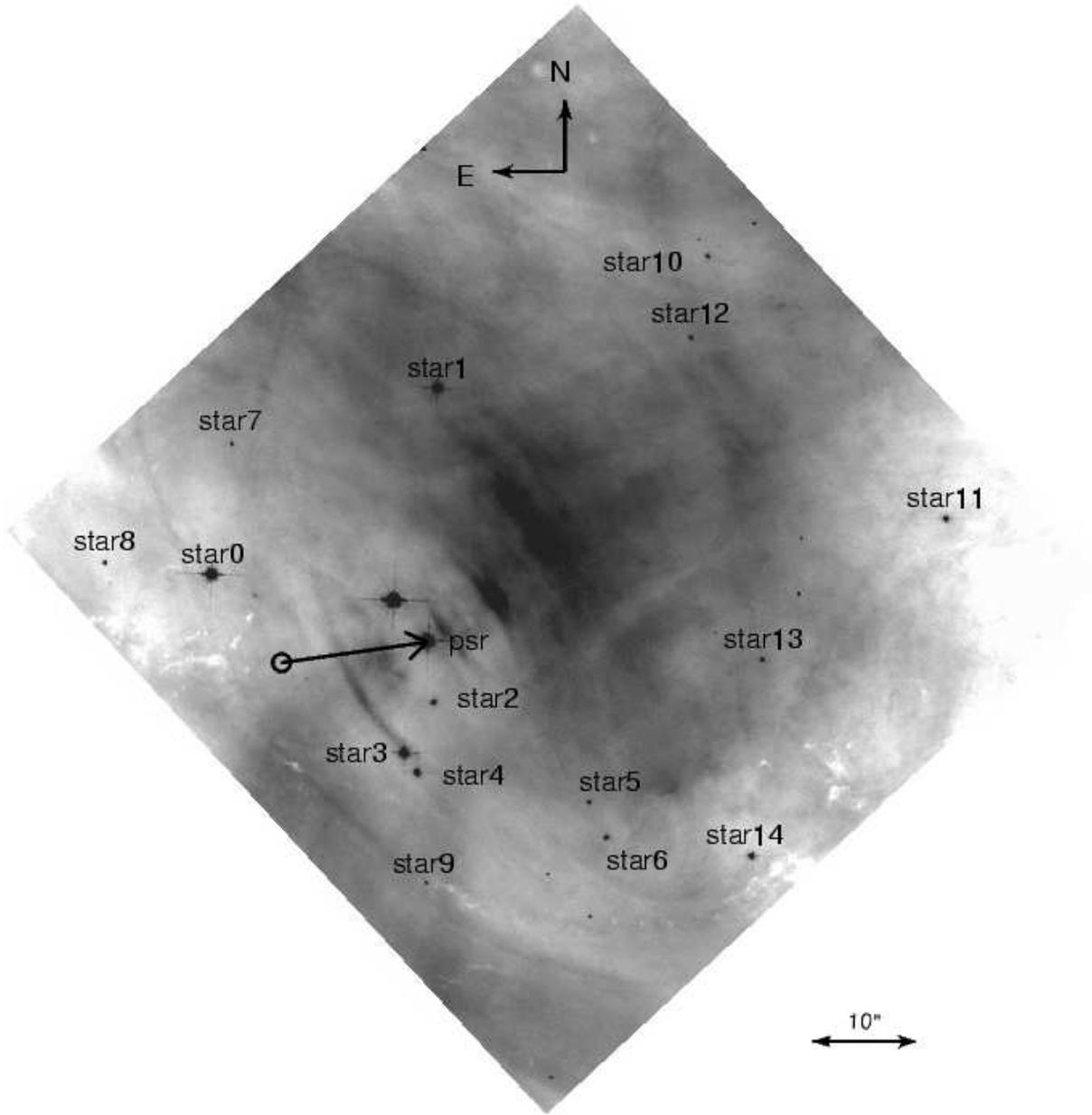}
\epsscale{0.8}
\caption{Co-added WF3 images of group~6 observations. The Crab pulsar and 15 reference stars are labeled. The arrow indicates the best-fit pulsar proper motion since birth, with the uncertainty in the birth site represented by the circle.\label{fig1}}
\end{figure}

As labeled in figure~\ref{fig1}, 15 stars with high signal-to-noise ratio (S/N) 
are chosen as reference points.  After cosmic ray removal, images for 
the same epoch are co-added.  Then we employed 2D Gaussian fit with 
the task \emph{fitpsf} for centering the objects.  We found that the 
pulsar position is not sensitive to the knot at the SE, since the 
former is much brighter. More generally, the nebular background does
not appear to significantly affect the astrometry of the bright stars
used here, as tested by variation in the extraction region and background
fitting algorithm.  On the other hand, the pulsar and a few of the 
brightest reference stars are saturated in these exposures. To test whether this
saturation affects our Gaussian fit astrometry, we attempted
fitting with saturated pixels masked using the Data Quality Files.
We found that the changes to individual star positions are much smaller
than the estimated error in our astrometry (\S~\ref{sec3.4}) and
the residuals in frame alignment are in fact slightly lower without masking.
Therefore, we did not mask out any pixels in the fit.

\subsection{Distortion Correction}

After the star coordinates are obtained, we must correct for geometric 
distortion before frame alignment.  The tasks \emph{metric} and 
\emph{wmosaic} in the \emph{STSDAS} package are commonly used for 
geometric distortion correction of the WFPC2. However, in addition to
the degradation resulting from re-sampling, the geometric corrections are
based on \citet{gil95}, with an rms residual of $\sim$10~mas. We clearly
need higher accuracy for the proper motion measurement.
The recent study of \citetalias{and03} gives a much better solution 
with accuracy better than 0.02 wide field (WF) pixels, i.e. 2~mas.
It is argued in \citet{kap02} that the chromatic variation in the distortion
is modest for filters at similar wavelengths, so we applied the \citetalias{and03}
solution for the F555W filter in our analysis.
The pixel coordinates are first corrected for the 34th row defect \citep{and99},
then the geometric distortion is removed using the third-order 
polynomials in \citetalias{and03}.

\subsection{Frame Alignment}

To proceed to relative astrometry, each frame is aligned to a common
sky position, measuring four fitting parameters (scale, x-, y-shift and rotation).
The reference star positions and proper motions (which even for typical
$\sim$10~km/s local Galactic velocity dispersion can be detectable over the full 
data span) are not, of course known {\it a priori}, and so are also determined
in the global fit to the data frames.  All parameters are fitted simultaneously 
by global $\chi^2$ minimization. If we attempt to use all F547M data (table~\ref{tab1}),
we find large reference star residuals for frames with different roll angles and
when the pulsar is present in different camera chips (figure~\ref{fig2}).
This is almost certainly due to residual uncorrected geometric distortion.
Thus in order to minimize the systematic errors, we have to further limit
the fit to a subset of the data with similar pointing parameters.
Groups 3 \& 6 are the place to start, since they have identical roll angles
and contain most of the frames. However, we also included group~1 in the
fit since it has a similar roll angle with groups~3 \& 6 and the very
long (6 year) time base can help overwhelm residual distortion errors from
the somewhat different chip placement of the pulsar in these frames.

\subsection{Sources of Error}
\label{sec3.4}
Before discussing the results, it is important to have a quantitative estimate 
of the uncertainties in the star astrometry. The main sources of error are 
star position measurement and residual geometric distortion.

The formal statistical errors in measuring the star centroids are essentially 
negligible ($<$1mas) since all the stars have high S/N.  Also, as noted above,
the results are quite insensitive to the background fitting method. Following 
\citetalias{car99}, we used different aperture sizes, but found that the best-fit positions 
only varied by 1~mas for apertures $>3\times$ the full width half-maximum (FWHM).
Of course the true errors in star centroiding are much larger than that suggested
by the centering algorithm $\chi^2$, since the PSF is substantially under-sampled 
by the WFPC2 pixels, giving large Poisson fluctuations at the peak. The true
astrometric uncertainty is in fact best measured by comparing star 
pixel coordinates among group~6 frames, since these (many) frames have an 
identical telescope pointing and thus residual image distortion cannot introduce 
position shifts between frames. This exercise gives an rms variance in each 
coordinate of 4.3~mas, our best estimate of the single star centroiding error.

Simple Gaussian fit could in principal dominate this error, so
we tried a different centering method using \emph{Tiny Tim} \citep{kri95}
simulated point-spread functions (PSFs).
The results are very similar to the Gaussian astrometry. For example 
comparison of obs~35 \& obs~37 shows
rms scattering of 3.2 \& 4.6 mas for Gaussian and PSF fit respectively.
This suggests that Gaussian fits do not significantly degrade the accuracy
of the centroiding of these bright stars (although they do not, of course
capture systematic PSF-dependent centroid shifts).

As mentioned, a simple fit to all of the F547M data shows that 
frames with different telescope pointings do not match well. This implies
that some residual geometric distortion remains even after applying the 
\citetalias{and03} solution. Of course, we do not expect to recover their full
accuracy since our dataset is taken with another filter. Also, their study used
an effective PSF (ePSF) determined from the actual exposures (of $\omega$ Centauri). Since
the centroiding depends on the PSF used, the distortion map is specific
to a particular ePSF scheme. Unfortunately, the Crab frames contain too few stars
to build up a local ePSF. 

To probe the level of astrometric error introduced
by the residual distortion, we compared star positions measured from
exposures with different roll angles.  As an example, we consider
obs~23 (from group~5) \& obs~37 (from group~6), which have a
$\sim$180$\arcdeg$ relative rotation.
Superposition of the two frames gives a rms variance of 10~mas for the 8
common reference stars. However, we noticed that some bright stars
exhibit large systematic residuals. In particular, star~0 \& star 3 are
shifted for 20 \& 15~mas respectively in the y-direction after alignment.
A comparison of the pulsar positions in the two exposures also yields a similar
amount of shift in this direction. Given that the brightness of the
pulsar is comparable to star~0, we speculate that the systematic
error is due to some fine structures in the PSFs of the bright stars.
To further characterize the shift, we performed the same analysis on
other frames between the two groups and found that the pulsar, star~0 \&
star~3 always have large systematic residuals in the same direction.
To check if a realistic model PSF fitting could improve the results
(as the position-dependent model PSF is plausibly closer to the true PSF),
we repeated the same process with centroiding using the \emph{Tiny Tim} simulated PSFs.
However, comparison of obs~23 with obs~37 gives a larger rms variance of 14~mas
and the systematic shifts in the pulsar, star~0 \& star~2 persist.
Clearly the detailed structures that cause systematic errors of the bright stars
are unmodeled by the simulated PSFs.  We conclude that \emph{Tiny Tim} PSF fit 
is not an improvement over Gaussian fit for our case.

On the other hand, comparison of frames in the same group shows small variance
in the star positions, although the offset in telescope pointings are
as large as 20 pixels within a group. This argues against saturation
or PSF under-sampling effects dominating the shifts and suggests that the
position-dependent PSF distortions are more moderate. Furthermore, \citetalias{and03}
found that their distortion solution is stable over a period of years and the
orbital effects such as jitter and breathing induce scale variations less than
$10^{-4}$ (i.e. $<$1 mas for our relative star positions). We therefore suspect
that the large roll angle-dependent residuals are due to non-axisymmetric structures
in the bright star PSFs, not modeled in the Gaussian or \emph{Tiny Tim} fits. However,
group~1 is at a very similar roll angle to groups ~3 \& 6 and should therefore
avoid the relative shift found in the bright-star PSFs. We thus included it in our
fit, even though it may contain residual position-dependent PSF distortions. 
We estimated this residual distortion by discarding star~0 in the frame comparison. 
This reduces the scatter to 7.6~mas when obs~23 is compared with obs~37.
After removing the astrometric error in quadrature,  this implies that residual
distortion contributes $\sim$6~mas to the uncertainty when the
\citetalias{and03} correction is applied to Gaussian-fitted star positions
under the F547M filter. To conclude, groups~1, 3~\&~6 have sufficiently similar
telescope pointings such that the residual distortion is minimal;
group~5 is at a very different roll angle and groups 2~\&~4 use different chips.
They thus have large, uncorrected residual distortions and we must exclude
them from the fit.

To remove this residual distortion, the full effective PSF approach
\citep{and00} would be required.  Unfortunately, the dearth of field stars makes it
difficult to build a useful PSF library. We therefore proceed with
Gaussian fits to seek the best current results. However, as this study
was being completed, we learned that Anderson and colleagues are attempting
a full ePSF study of the Crab field and we expect that this can
provide additional astrometric accuracy.

\section{Results}
\begin{figure}[!ht]
\plotone{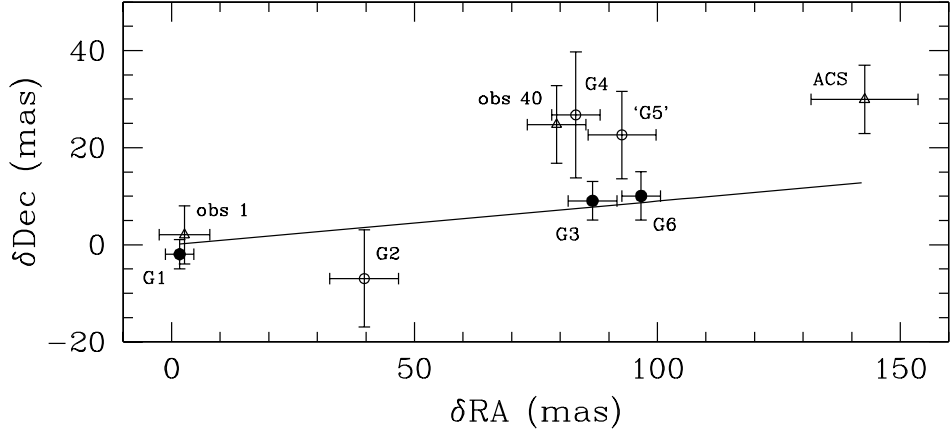}
\caption{Position of the Crab pulsar relative to obs~2.
The reference star positions and proper motions are fitted from groups~1,
3 \& 6 while individual frames are aligned with obs~2.
Filled circles: group~1, group~3 \& group~6; 
open circles: group~2, group~4~\&~`group~5' (the mean of obs~21, 23 \& 25 
is shown); open triangles: obs~1, 40~\&~41.
The best-fit pulsar proper motion (solid line) is fitted to the filled-circle
data only. The other epochs have large unmodeled geometric distortions that affect
the relative pulsar astrometry, as shown (see text). Note, however, that these
epochs do confirm the general proper motion trend.
\label{fig2}}
\end{figure}

Figure~\ref{fig2} shows the pulsar positions at different epochs, with the
star positions and proper motions fixed from the groups~1, 3 \& 6 fit
and frame registration determined by standard transformations of individual epochs.
The uncertainties are given by rms scatterings of the reference 
star coordinates in the corresponding epoch. In each case we re-scale the
errors accounting for the number of degrees of freedom absorbed by the
fitted parameters.  Since there are only two independent frames in group~1, 
a substantial fraction of the position uncertainty is absorbed by the
reference star proper motion terms in the fit. In addition, for group~1
the pulsar position on the chip differs substantially (although the observation
roll angle is similar) so residual geometrical distortions almost certainly
affect the pulsar astrometry, as described above. Thus the re-scaled statistical errors
(solid error flags) are likely significantly too small. So 
we add, in quadrature, the 4.3~mas centroiding error and the 6~mas
residual distortion error estimated above for a more realistic 7.4~mas
error estimate (dashed error flags). This gives a more conservative
error in the proper motion fit.

Linear regression is employed to determine the pulsar's proper motion. The 
best-fit result
\[ \mu''_\alpha = -15.0\pm 0.8 \;\mathrm{mas/yr,} \;\; 
\mu_\delta = 1.3\pm 0.8\;\mathrm{mas/yr} \]
is plotted by the solid line in the graph.
While $\mu''_\alpha$ is consistent with previous studies, $\mu_\delta$ is about 5 times smaller,
resulting in a significantly different position angle for the proper motion.
(For reference, we obtained $\mu''_\alpha=-15.8\pm0.3\;\mathrm{mas/yr,} \; \mu_\delta=1.4\pm0.3\;\mathrm{mas/yr}$ if the statistical errors of group~1 are used).

\begin{figure}[!ht]
\plotone{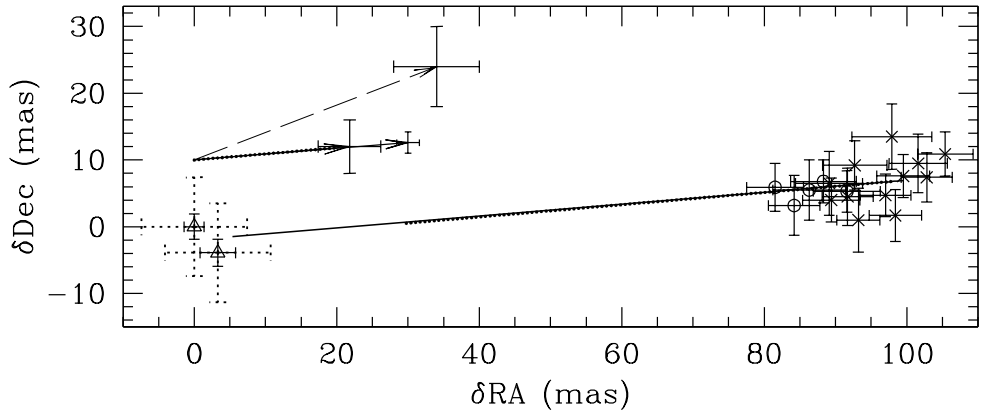}
\caption{Best-fit proper motion of the Crab pulsar.
The pulsar's positions relative to obs~2 are plotted for different epochs 
(open triangle: group~1; open circle: group~3; cross: group~6).
The solid and dotted error flags of group~1 are statistical and 
systematic errors as discussed in the text.
The best-fit proper motions for the group~1, 3, 6- and group~3, 6- fits are 
shown by the solid and dotted lines respectively;
the length represents the motion from obs~2 (MJD 49723) to obs~39 (MJD 52016).
The dashed, solid and dotted arrows indicate the best-fit pulsar motions in two 
years for \citetalias{car99} and our two fits respectively, all without Galactic 
rotation and solar motion corrections.\label{fig3}}

\end{figure}

We also checked the fit with only group~3~\&~6 data, since residual
distortion cannot affect our astrometry in these images. For this restricted
data set, the proper motion
\[ \mu''_\alpha = -10.9\pm 2.2 \;\mathrm{mas/yr,} \;\; 
\mu_\delta = 1.0\pm 2.0\;\mathrm{mas/yr}\]
is shown by the dotted line in the figure~\ref{fig3}. The errors are of course larger with
the shorter time base, but systematic errors {\it cannot} contribute significantly.
Note that $\mu''_\alpha$ differs by $\sim$1.7$\sigma$. $\mu_\delta$ is
not significantly changed, and still differs substantially from
previous estimates.

The best-fit pulsar proper motions (two years' position offsets) with the
associated uncertainties are shown by dashed, solid and dotted arrows in 
the graph for \citetalias{car99} and the two above cases, respectively.
The proper motion directions in our two fits are in excellent agreement,
with a posiiton angle $275\arcdeg\pm3\arcdeg$. For comparison, the \citetalias{car99} 
best-fit value (which is similarly uncorrected to the local standard of rest) 
lies 5.7$\sigma$ away, although given their much larger errors the measurements
only disagree at the 1.6$\sigma$ level.

To convert the proper motion to its local standard of rest,
Galactic rotation and solar peculiar motion corrections must be applied.
Here a flat rotation curve is assumed, with $\Omega_0=220$ km/s and 
$R_0=8.5$ kpc \citep{fic89} and the solar constants are taken to be 
$10\pm0.36$, $5.25\pm0.62$ \& $7.17\pm0.38$ km/s \citep{deh98}.
Applied to our group~1, 3 \& 6 fit, this gives our final result of
\[\mu''_\alpha = -14.7\pm0.8\;\mathrm{mas/yr},\;\;  \mu_\delta = 2.0\pm0.8\;\mathrm{mas/yr} \]
or
\[\mu_\ast = 14.9\pm0.7\;\mathrm{mas/yr}\;\;  \mbox{at PA } 278\arcdeg\pm3\arcdeg. \]
At the nominal distance of 2 kpc, this gives the pulsar's space velocity as $140\pm8$ km/s .
For reference, the group~3 \& 6 solution has $\mu_\ast=10.8\pm2\;\mathrm{mas/yr,}$ at PA $279\arcdeg\pm10\arcdeg$ in its local standard of rest.

\section{Discussion}

Our study, using WFPC2 observations spanning $>$6 years, improved geometric 
distortion corrections and a more realistic assessment of errors, suggests a 
new value for the Crab pulsar's proper motion differing from that of 
previous studies. We have not been able to reproduce the estimate of 
\citetalias{car99}. We believe that the errors were substantially underestimated in their 
proper motion study which used only four stars in three data frames spanning 1.9 years.
We must conclude that our new position angle measurement $278\arcdeg\pm3\arcdeg$ 
supersedes this earlier estimate. 

This is important, since the apparent alignment of the Crab pulsar's proper motion
and spin axis (as measured from the pulsar wind nebula) has significantly driven
thinking about the linear momentum-angular momentum correlations expected
in different models of the birth kick. Indeed, as the index case, the Crab pulsar has
suggested that many pulsars may have spins and kicks aligned at birth. A 
simplified interpretation \citep{spr98,rom05} suggests that during core collapse, 
the momentum imparted by the kick is rotationally averaged, resulting in 
alignment between velocity and spin vectors.  Hence the alignment angle 
is an important observational parameter that places constraints on the characteristic 
timescale of the kick.  \citet{ng04} derived the Crab pulsar's spin axis at PA 
$124\fdg0\pm0\fdg1$ (equivalent to $304\fdg0 \pm 0\fdg1$) by pulsar wind torus fitting.
Comparing with the best-fit proper motion leads to a misalignment of $26\arcdeg\pm3\arcdeg$,
which is substantially larger than the previously suggested value.
This is also greater than that of several other pulsars, including the Vela 
pulsar \citep{ng04}.
Note that the measured angle is the projection on the sky plane of the true 3D 
alignment angle.  The latter cannot, of course, be deduced unless the
pulsar's radial velocity is known.  However, using the 3D orientation of the 
nebula's symmetry axis \citep{ng04}, simple geometry shows that for a random pulsar 
radial velocity between -500 km/s to 500 km/s,
there is only 8\% chance to obtain a 3D alignment angle {\emph smaller} than the 
projected value. In fact, the 3D mis-alignment is always greater than $23\arcdeg$
for any value of the radial velocity.
The large alignment angle relaxes the constrains on pulsar kick during core collapse.
Since the initial spin period for the Crab pulsar is about 19~ms,
our result suggests that a short kick timescale is allowed, possibly much 
less than 1s. Of course, a more detailed treatment of the spin induced by the 
off-center
kick can complicate the interpretation \citep[][Ng and Romani, in preparation]{spr98}.

Our analysis is clearly not the last word on the important question of the Crab pulsar's
proper motion.  Improved astrometry, with more accurate distortion
correction, and an improved time base employing newly scheduled ACS images should
allow the use of all the HST images and
further reduce the proper motion errors. Perhaps a final measurement with the
WFPC2, with the pointing parameters of group~1 would also be useful, ensuring the
minimal sensitivity to systematic errors. Pursuit of such measurement is quite
important since our result, at a minimum, calls in to question the widely accepted alignment of the Crab pulsar's spin and velocity. 

\acknowledgments

This work was supported in part by NASA grant NAG5-13344.

\end{document}